# Measuring religious morality using very limited poll responses: Implementing "big-data analytics" to small data

Alvin Vista

## Abstract

Opinion polls remain among the most efficient and widespread methods to capture psycho-social data at large scales. However, there are limitations on the logistics and structure of opinion polls that restrict the amount and type of information that can be collected. As a consequence, data from opinion polls are often reported in simple percentages and analyzed non-parametrically. In this paper, response data on just four questions from a national opinion poll were used to demonstrate that a parametric scale can be constructed using item response modeling approaches. Developing a parametric scale yields interval-level measures which are more useful than the strictly ordinal-level measures obtained from Likert-type scales common in opinion polls. The metric that was developed in this paper, a measure of religious morality, can be processed and used in a wider range of statistical analyses compared to conventional approaches of simply reporting percentages at item-level. Finally, this paper reports the item parameters so that researchers can adopt these items to future instruments and place their own results on the same scale, thereby allowing responses from future samples to be compared to the results from the representative data in this paper.

# 1 Introduction

## 1.1 Challenges in public opinion polls

Opinion polls are one of the main sources of information for the general public and for various stakeholders across a wide range of fields but more prominently in politics, economics, marketing and social advocacy (e.g., surveys on loneliness, happiness, well-being). Opinion polls are among the most efficient methods to gather mass data on psycho-social constructs[1]. Compared to formal assessments, psychological evaluations, and standardized tests, opinion polls are faster and cheaper. These advantages, however, are countered by the low response rates of opinion polls, especially for telephone surveys. The response rates of telephone surveys can be as low as 9% and online polls, while more utilized as a method, have even lower response rates (Keeter, Hatley, Kennedy, & Lau, 2017). Even high-effort methods, such as in-person surveys and those that include monetary incentives or longer field periods, the response rates have also been declining over the years and more recently stands at around 22% (Keeter, Kennedy, Dimock, Best, & Craighill, 2006). Although low response rates are not an indicator of quality and can produce comparable results as high-response methods, the low response rates have implications for costs as more people would need to be contacted to obtain a target sample size (Keeter et al., 2006). The nature of opinion polls also restricts the *amount* and *type* of information that can be collected. No one wants to go through a two-hour telephone survey and, by its nature, all responses to an opinion poll are essentially self-reports. The main challenge given these constraints is how to extract more reliable information from even very restricted opinion poll data.

This paper discusses the validity and interpretability issues in the traditional/conventional approaches to dealing with categorical response data, while also demonstrating the potential of modern measurement approaches to address these issues and challenges. The main purpose of this paper is to compare conventional vs modern measurement approaches to develop useful metrics based on very limited survey responses. These metrics are essential if we need to analyze the results in more comprehensive manner because it will allow for more sophisticated quantitative methods to be applied to interval-level metrics than to the original categorical responses.

---

[1] In psychological terminology, "constructs" are concepts that have meanings "constructed" from theory rather than from empirical sources. That is, unlike empirical entities with tangible and measurable properties, constructs have no inherent meaning independent of their theoretical conceptualization.

For the purposes of this paper, we are using a national telephone survey on political issues to demonstrate that we can extract a limited subset of the response data to develop a metric on "religious morality". Large-scale national opinion polls commonly gather information on multiple issues and therefore there is potential to extract data on various psycho-social constructs. However, these subsets of any particular opinion poll (which is itself limited in scope) are very limited, and hence the need to implement methods to maximize the amount of information that we can extract and process.

## 1.2 Conventional approaches

The most common approach in reporting categorical response data is through frequency counts, for example as shown in Table 1. A related approach is to report percentages, which is popular in the media. We often hear of reporters saying that, "X% of people disapproves of this particular person or policy". There is nothing wrong with this usage of opinion data, but it is very limited and specific to a particular survey question. This means that we cannot infer beyond what information is available in the proportions of responses to a category, nor can we compare the numbers across the survey questions (i.e., a 40% agreement on a question is not interpretable in the same way as a 40% agreement on another question).

Conventional options for analytical approaches are rather limited to non-parametric statistical analysis in the strict sense. However, it is not uncommon for researchers to treat Likert-type categorical responses as if these are on an interval scale and apply parametric analysis on the data.

Treating ordered categorical data as an interval measure rest on the assumption that the distance between adjacent categories are uniform (e.g., the difference between "strongly agree" and "agree" is quantitatively the same as between "disagree" and "strongly disagree"), which in turn requires that the qualitatively described categories mean the same for different people endorsing them, and assumption that has long been under contention (see Cronbach, 1946; Jones & Thurstone, 1955).

Likert (1932) himself considered the summation of item scores (i.e., total of the coded responses) as a scale, although he did not consider the item-level codes themselves as a scale. Likert's position adds to the confusion because it is mathematically unclear as to how codes that are not considered scales can sum up to a scale. Nevertheless, the popularity of Likert-type scales is undeniable, and the use of summated scores and treating them as an interval scale is ubiquitous across the social sciences (Cummins, 1997). This is a controversial issue and the debate as to the extent that the assumption of an interval scale is defensible continues.

| Party affiliation | Q40c: Do you personally believe that having an abortion is morally acceptable, morally wrong, or is it not a moral issue? | | | | |
| --- | --- | --- | --- | --- | --- |
| | Depends on situation | Don't know/Refused | Morally acceptable | Morally wrong | Not a moral issue |
| Democrat | 51 | 18 | 90 | 162 | 157 |
| Independent | 39 | 21 | 65 | 226 | 155 |
| No preference | 7 | 4 | 6 | 19 | 4 |
| Other party | 4 | 3 | 5 | 8 | 6 |
| Republican | 28 | 16 | 34 | 298 | 68 |

*Table 1*

## 1.3 Item response modelling approaches

Item response modeling (IRM) is part of a family of factor analytic approaches towards measurement of latent traits based on item response theory (IRT) of measurement[2]. Under this approach, these item response models yield interval-level measures which are more useful than the strictly ordinal-level measures obtained from Likert-type scales (Goldstein & Hersen, 1984). The IRM approach involves models that have a wide range of complexity, ranging from models that are designed for dichotomous responses (e.g., 0-1, yes-no) to those that handle multiple categories and types of responses. Taking into account that responses to Likert-type questions are not in a true interval scale, we used item response models that are designed for categorical responses. Three item response models were used in this paper, each have different design parameters which will be discussed in detail in the Methods section.

# 2 Methods

## 2.1 Data

The dataset[3] came from the February 2012 political survey conducted by the Pew Research Center (2012). This dataset contains a national sample of *N*=1501 respondents (all adults and living in 50 US states and the District of Columbia) to telephone interviews conducted using random digit dialing (for more details on the survey methods, see Pew Research Center, 2018). Four of the survey questions were used for this paper, described in Table 2. To make the ordered categories uniform across the "morality" and the "religiousness" questions, the response categories were reordered such that larger values are interpreted consistently across these two main scale components. While this is technically just an arbitrary ordering, we have included an item response model that does **not** take this *a priori* ordering into account to check if our assumptions are reasonable.

---

[2] For a more comprehensive dive on this rich topic, Embretson & Reise (2013) and Wilson (2004) provide an excellent course on the fundamentals of IRT.

[3] This dataset can be downloaded at http://www.people-press.org/dataset/february-2012-political-survey/

| Variable | Question | Response options | | Recoded ordered categories | |
|---|---|---|---|---|---|
| Q40a | Do you personally believe that **using contraceptives** is morally acceptable, morally wrong, or is it not a moral issue? | 1 | Morally acceptable | 1 | Morally acceptable |
| | | 2 | Morally wrong | 2 | Not a moral issue |
| | | 3 | Not a moral issue | 3 | Depend on situation |
| | | 4 | Depend on situation | 4 | Morally wrong |
| Q40b | Do you personally believe that **getting a divorce** is morally acceptable, morally wrong, or is it not a moral issue? | 9 | Don't know/Refused | NA | Don't know/Refused |
| Q40c | Do you personally believe that **having an abortion** is morally acceptable, morally wrong, or is it not a moral issue? | | | | |
| ATTEND | Aside from weddings and funerals, how often do you attend religious services… more than once a week, once a week, once or twice a month, a few times a year, seldom, or never? | 1 | More than once a week | 6 | More than once a week |
| | | 2 | Once a week | 5 | Once a week |
| | | 3 | Once or twice a month | 4 | Once or twice a month |
| | | 4 | A few times a year | 3 | A few times a year |
| | | 5 | Seldom | 2 | Seldom |
| | | 6 | Never | 1 | Never |
| | | 9 | Don't know/Refused | NA | Don't know/Refused |

*Table 2*

Cases with non-response on all of the "morality" questions (Q40a to Q40c) were not included in this paper, yielding a final sample size of *N*=1494 for the analyses in this paper. Selected demographic variables from the dataset were used as grouping variables in the analyses. Table 3 reports the grouping variables and the categories within each group, as well as unweighted and weighted sample sizes based on the sampling weights provided in the dataset. The weighted sample size was used in computing the standard errors and the confidence intervals in the analyses. Additional details on the sampling methodology and the definitions of these demographics variables can be obtained from the original survey (Pew Research Center, 2012) and the Center's published methodology references (Pew Research Center, 2018).

| **Demographic Variables** | **Unweighted N** | **Weighted N** |
|---|---|---|
| Party affiliation | | |
| Independent | 506 | 1683.00 |
| Democrat | 478 | 1480.46 |
| Republican | 444 | 1220.54 |
| No preference | 40 | 160.23 |
| Other | 26 | 107.88 |
| Gender | | |
| Female | 907 | 2797.88 |
| Male | 533 | 1564.08 |
| Missing data | 54 | 290.15 |
| Race | | |
| Black or African-American | 626 | 1796.08 |
| White | 550 | 1783.27 |
| Hispanic | 144 | 425.54 |
| Some other race | 96 | 276.15 |
| Asian or Asian-American | 24 | 80.92 |
| Missing data | 54 | 290.15 |
| Geographical area | | |
| Suburban | 681 | 2043.92 |
| Urban | 454 | 1518.23 |
| Rural | 284 | 786.31 |
| Missing data | 75 | 303.65 |
| Region | | |
| South | 530 | 1665.00 |
| Midwest | 368 | 1057.81 |
| West | 365 | 1091.46 |
| Northeast | 231 | 837.85 |
| Political views/ideology | | |
| Moderate | 542 | 1708.31 |
| Conservative | 438 | 1270.88 |
| Liberal | 241 | 763.65 |
| Very conservative | 109 | 300.08 |
| Very liberal | 93 | 334.92 |
| Don't know/Refused | 71 | 274.27 |
| Family income | | |
| Less than $10,000 | 100 | 452.35 |
| 10 to under $20,000 | 142 | 507.77 |
| 20 to under $30,000 | 174 | 635.58 |
| 30 to under $40,000 | 147 | 456.42 |
| 40 to under $50,000 | 140 | 396.08 |

| 50 to under $75,000 | 199 | 570.50 |
| 75 to under $100,000 | 157 | 457.42 |
| 100 to under $150,000 | 165 | 430.42 |
| $150,000 or more | 113 | 276.65 |
| Don't know/Refused | 157 | 468.92 |

*Table 3*

## 2.2 Scaling and analysis

### 2.2.1 Modeling the classical composite scale

A common, if somewhat controversial, method used for analyzing ordered categorical data is simply using the mean of the composite score across the response data. In situations where the categories are not equally spaced, the responses can be standardized or converted into a z-score.

We used two standardization approaches here for comparative purposes. The first is to standardize the composite score after it has been calculated. Because standardization is a linear transformation, it does not matter if the conversion is done on the raw total score or on the mean score. The mean z-score in this approach is denoted as **z_score.b**.

$$\text{z\_score.b} = \frac{x_n - \mu_N}{\sigma_N}$$

where $x_n$ is the composite score for person $n$, and $\mu_N$ $\sigma_N$ represent the overall mean and standard deviation respectively.

The second approach is to standardize each response first and then computing the average of the z-scores across all $Q$ items. The mean z-score in this approach is denoted as **z_score.w**.

$$\text{z\_score.w} = \frac{\sum_1^Q \frac{x_{ni} - \mu_{Ni}}{\sigma_{Ni}}}{Q}$$

where $x_{ni}$ is the score for person $n$ on item $i$, and $\mu_{Ni}$ $\sigma_{Ni}$ represent the overall mean and standard deviation for all respondents on item $i$ respectively.

The difference between these two approaches is that the final mean scores in the first approach are comparatively more skewed towards the items with the largest category value. For example, if a questionnaire has 9 items with a 5-point scale, and one item with a 10-point scale, the totals and hence the means will be more heavily affected by the 10-point item. Whereas in the second approach, because all responses were standardized prior to aggregating, all items have the same contribution towards the aggregate score.

### 2.2.2 Modeling the GRM scale

Using a graded response model (GRM; Samejima, 1968), also known as the ordinal response model, the responses in the 4 questionnaire items were treated as ordered categories similar to a rating scale or Likert-type items. This means that the intervals between categories are not interpreted as numeric scores but rather as nominal categories that have a specific order. Under the GRM, for item $i$ with $m$ possible response categories coded as $x$ the probability of a person $n$ with latent ability $\theta_n$ responding with category $x$ <u>or greater</u> is expressed below:

$$P_{ix}^*(\theta_n) = P(X_i \geq C_{xi}|\theta_n) = \frac{e^{a_i(\theta_n - d_{ix})}}{1 + e^{a_i(\theta_n - d_{ix})}}$$

where $a_i$ and $d_i$ represent the discrimination and difficulty parameters of item $i$ respectively. In the context of subjective responses such as in opinion polls, the difficulty parameter can be interpreted as the location of the particular question statement on the opinion continuum (e.g., level of agreement).

This is the *cumulative category response function* where the *boundary* of response probabilities is defined as follows:

$P(X_i \geq C_{1i}|\theta_n)$ = 1, responding with the lowest category or greater

$P(X_i \geq C_{mi}|\theta_n)$ = 0, responding with the highest category or greater

As such, the probability of a person responding <u>exactly</u> with a given category is:

$$P_{ix}(\theta_n) = P_{ix}^*(\theta_n) - P_{ix+1}^*(\theta_n)$$

$$P_{ix}(\theta_n) = P(X_i = C_{xi}|\theta_n) = \frac{e^{-a_i(\theta_n - d_{ix+1})} - e^{-a_i(\theta_n - d_{ix})}}{[1 + e^{-a_i(\theta_n - d_{ix})}][1 + e^{-a_i(\theta_n - d_{ix+1})}]}$$

### 2.2.3 Modeling the GGUM scale

The generalized graded unfolding model (GGUM; Roberts, 2008) operates on the basis that subjective responses are not strictly cumulative even if they are ordered in nature. The concept behind the GGUM is that item responses unfold from the position that centers on an individual's location on the scale (Roberts, Donoghue, & Laughlin, 2000). This model is more appropriate for Likert-type items where the item-format measures extent of agreement and there is no strict requirement for the graded categories to be cumulative (e.g., in GRM). Under this model, the responses can be ordered where the direction of the ordering is symmetric from a category that is indicative of one's position on a continuum. For example, if someone responds "Agree" on a question, the person might be coming from both sides of the continuum and the response probabilities are equal from either side. In other words, the response probabilities "unfold" from one's underlying position on the continuum where for every response category, an individual has two possible subjective response: "endorse the item from the left end of the continuum" or "endorse the item from the right end of the continuum".

If we denote $C$ as the number of observable response categories minus one (i.e., excluding the actual observed response, or the person's location on the continuum), the unfolding mechanism can be thought of as implying that there are two "subjective" responses on either side of the observed response. If we denote $M$ as the number of "subjective" responses, then $M = 2*C + 1$. The responses can then be reparametrized as distance ($z$) from the person's location, where $z = 0, ..., C$, and $C$ by definition above also represents the maximum level of agreement corresponding to the person's location (the point where $\theta_n = d_i$). The GGUM defines the probability of a person $n$ with latent ability $\theta_n$ responding with $z$ on item $i$ as:

$$P(Z_i = z|\theta_n) = \frac{f(z)-f(M-z)}{\sum_{w=0}^{C}[f(z)-f(M-z)]}, \text{ such that}$$

$$f(w) = \exp\left\{a_i\left[w(\theta_n - d_{ik}) - \sum_{k=0}^{w}\tau_{ik}\right]\right\}$$

where $\tau_{ik}$ represents the threshold parameter for the $k$th response category relative to the location of the $i$th item. This parameter is constrained to 0 at the neutral position, such that $\tau_{i(C+1)} = 0$. And because the probabilities for opposing categories are symmetric from the neutral position, $\tau_{iz} = -\tau_{i(M-z+1)}$ for $z \neq 0$.

The main difference between the GRM and GGUM is that GRM is cumulative in the sense that an endorsement of a category implies endorsement of all categories that are ordered lower than it. Endorsement of a category under GGUM however is associated with a subjective position that could come from opposing poles with respect to the statement being endorsed. In a pure rating scale format, where the response indicates a level of agreement towards a statement, the GGUM might be more appropriate. However, the questions that were used for this study is a mix of rating and categorically ordered formats.

### 2.2.4 Modeling the NRM scale

The final model we include in this comparison does not constrain the response categories with an *a priori* ordering. Although from the layperson's perspective it would be reasonable to interpret the categories as ordered, it is not strict requirement and models that do not have this constraint is useful to test whether the a priori ordering set by the questionnaire developer is valid in the first place. For example, although we recoded a response of "Not a moral issue" as lower on the scale compared to "Depend on situation", this interpretation might not be universal.

For comparative purposes, we included an item response model that does not have a prior assumption of rank order among the categorical response. In the nominal response model (NRM; Bock, 1972), the probability of a person $n$ with latent ability $\theta_n$ responding with category $k$ on item $i$ is given as:

$$P(X_i = k_i | \theta_n) = \frac{e^{a_{ik}(\theta_n - d_{ik})}}{\sum_{h=1}^{m} e^{a_{ik}(\theta_n - d_{ih})}}$$

where $k_i$ represents the $k$th option, $k \; \varepsilon \; [1,2, …,m]$ out of $m$ options. Similar to the GRM, $a_{ik}$ and $d_{ik}$ represent the discrimination and difficulty parameters of item $i$ respectively for each category $k$.

### 2.2.5 Locating individual position on the scale

All item response models were fitted to the data using the **mirt** package (Chalmers, 2012), where the item parameters were estimated using an expectation-maximization algorithm (Bock & Aitkin, 1981). Once the item parameters were estimated, point estimates of the respondents' location on the measurement scale are computed. This person parameter ($\theta_n$) is based on the likelihood of the person's response vector given the item parameters, and estimated using either maximum likelihood or Bayesian estimators. The latter type, specifically the modal a posteriori (MAP) estimator, was used in this paper to locate the respondents on the respective scales. MAP estimation is described more fully by Embretson and Reise (2013). Estimating $\theta_n$ is as follows:

$$\hat{\theta}_n = Y(\theta_n | \boldsymbol{\xi}, \mathbf{U}_n)$$

where *Y* is the expected θ given ξ as the matrix of item parameters (e.g., $a_i$, $d_i$) and $\mathbf{U}_n$ as the vector of responses ($U_{1n}$, $U_{2n}$, ... $U_{in}$) to each item (1,2,... *i*) by person *n*.

The resulting scales for both z-scores and MAP estimates have the same metric characteristics (*M*=0, *SD*=1) so they are broadly comparable. The scales are centered at 0, which means that the overall average "religious morality" scale score is 0, and higher values indicate stronger religious morality. The scale is probabilistic, as opposed to a physical scale where values represent real magnitudes, which means that the values indicate relative probabilities. The units of the scale are in logits (log of the odds). To illustrate this scale, Figure 1 shows the location of four individuals along a continuum, a scale of some construct (e.g., "morality"). Because the construct itself is not physical, the scale is arbitrary in nature and can only show relative relationships (i.e., relative to the center of the scale). In this case, the values indicate associations to the scale relative to someone with a scale score of 0, with odds defined as: $O_{\text{MAP}} = \frac{p}{1-p}$. As such, Person A has an odds of 0.082 [$O_{\text{MAP}}$ = .07586/(1-.07586)] being associated with the construct relative to the scale center, while Person D has an odds of 33.115.

The characteristic of an interval-level scale means than two individuals with MAP estimates that differ by X will have the same relative probability of being associated with a location on the scale, or the same odds-ratio (*OR*) no matter where they are on the scale. That is, the *OR* will be the same between Person A(MAP=-2.5) vs Person B(MAP=-1) as between Person C(MAP=2) vs Person D(MAP=3.5). The logs of these differences will also be the same, log(.22)=-1.5, which reflect the same differences between the pairs A-B and C-D.

$$OR_{(A/B)} = \frac{O_{\text{MAP(A)}}}{O_{\text{MAP(B)}}} = \frac{\frac{p_A}{1-p_A}}{\frac{p_B}{1-p_B}} = \frac{0.082}{0.368} = 0.22$$

$$OR_{(C/D)} = \frac{O_{\text{MAP(C)}}}{O_{\text{MAP(D)}}} = \frac{\frac{p_C}{1-p_C}}{\frac{p_D}{1-p_D}} = \frac{7.389}{33.115} = 0.22$$

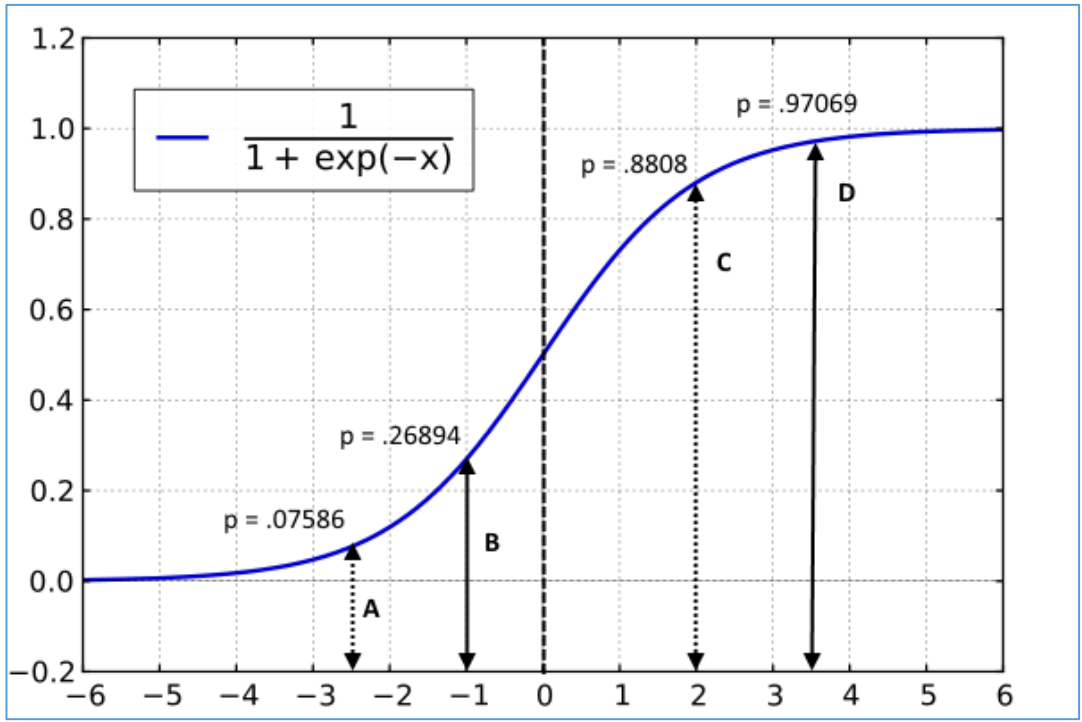

*Figure 1*

Finally, summary statistics were computed for the subgroups that were chosen previously. All analyses were conducted within the **R** environment (CRAN, 2017).

## 3 Results

### 3.1 Comparison between the different standardization methods in the classical approach

As previously discussed, **z-score.b** estimate is more skewed towards the items with the largest category value compared to the **z-score.w** estimate. In this instance, the "religiosity" question has more impact on the aggregate score because its code ranges from 1 to 6 whereas the rest of the questions are on a 1-4 range. Checking whether this effect is statistically significant, we can see that using the **z-score.b** estimate produces a statistically significantly different mean score for some subgroups. Figure 2 shows that the mean scale score for Republicans is overestimated compared to the mean scale score using **z-score.w** estimates. This overestimate is statistically significant at a confidence level of 95%. The effect is even more pronounced in the mean scale scores across subgroups based on political ideology (Figure 3), where statistically significant biases are observed for the mean scale scores of 4 out of 6 subgroups. More importantly, this bias affects the inferences we make. Figure 3 shows that if **z-score.b** estimate is used, it would support the inference that the very liberal group have a statistically *lower* average religious morality scale score than the liberal group, an inference that would not be supported if the **z-score.w** estimate is used.

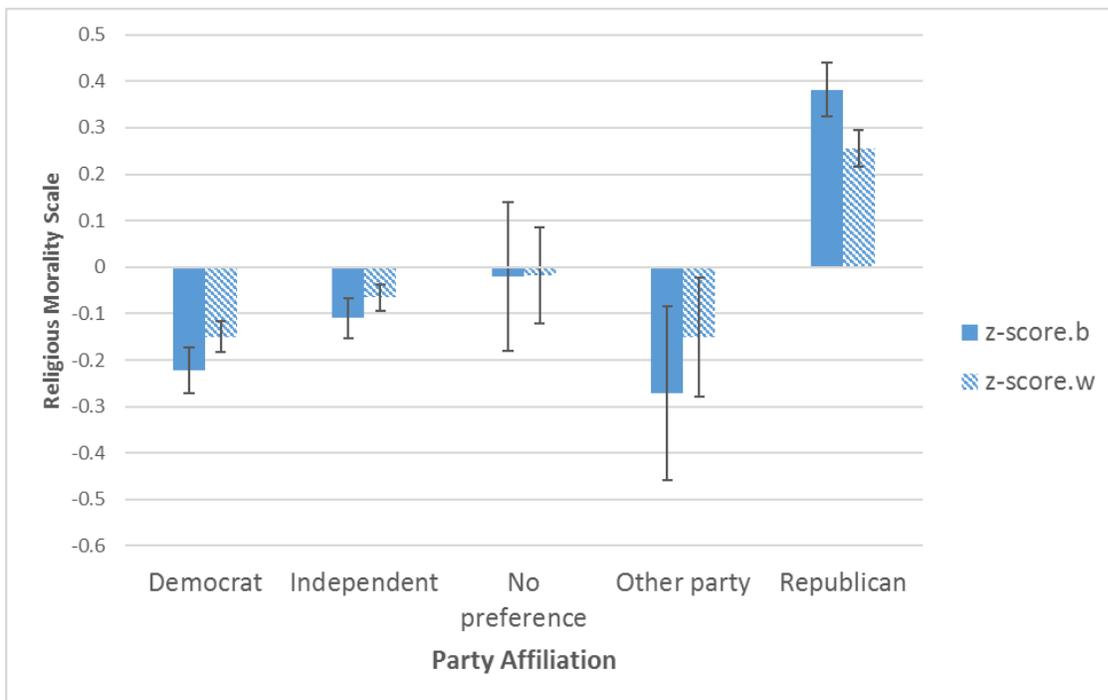

*Figure 2*

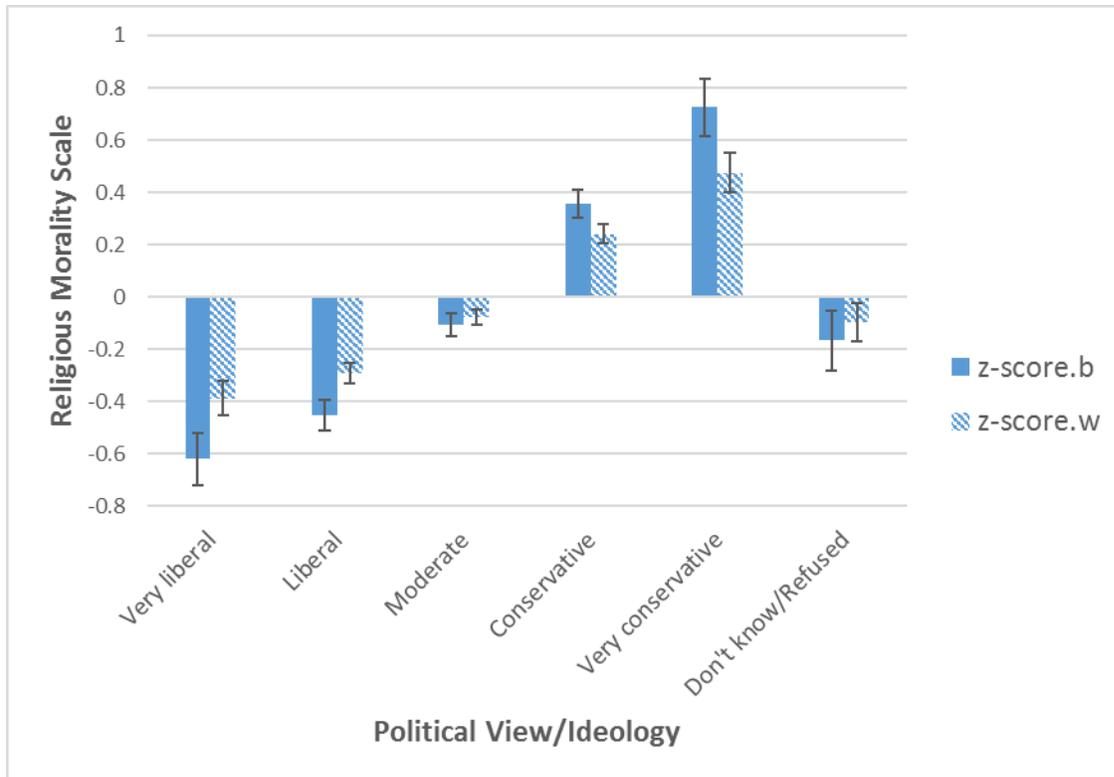

*Figure 3*

Notwithstanding the controversy around the use of averages in Likert-type scales, this comparison of the estimates from the naïve approach shows that standardizing at item level prior to computing the summated scale (and thereafter the average for such scale) is perhaps the lesser evil. The caveat in these comparisons, however, is that if the assumption fails that Likert-type scales can be treated as interval scales, then these comparisons will not be mathematically defensible because the standard errors (which are the bases for the confidence intervals) will not be interpretable.

## 3.2   Comparison between the various IRM-based scales

In contrast to the naïve approach in dealing with categorical scores, the IRM-based scales are proper interval scales and therefore the distances are uniform. The latent estimates are probabilistically derived and the computation of standard errors is mathematically defensible. The latent estimates are dependent on the particular item response model that is used to obtain them, but the scales have the same characteristics and are comparable with each other. Because the scales are interval measures, the averages for any group or subgroup are also comparable to other groups. For example, the average scores by party affiliation (Figure 4) can be compared directly with the average scores by political view (Figure 5) or by geographic region (Figure 6).

If we need to compare the models, the Akaike information criterion (AIC) and Bayesian information criterion (BIC) can be used where lower values are interpreted as better fitting models (Raftery, 1995). Table 4 shows that the NRM is more preferable, although the estimates from each of the three models are in close agreement and often do not have statistically significant differences from each other. The only notable difference is in the estimates for the Independent (see Figure 4) and Moderate (see Figure 5) subgroups, where the average scale score based on the GGUM estimate is significantly different from the other two models.

| Model | Model AIC | Model BIC |
|---|---|---|
| GRM | 14527.03 | 14622.60 |
| GGUM | 17211.51 | 17328.32 |
| NRM | 14199.43 | 14348.09 |

*Table 4*

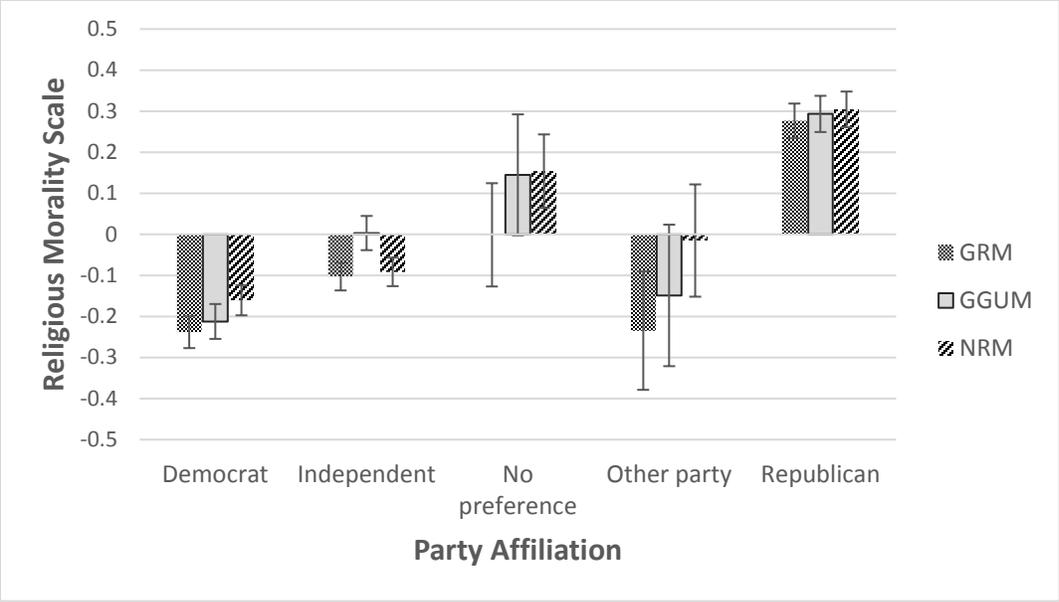

*Figure 4*

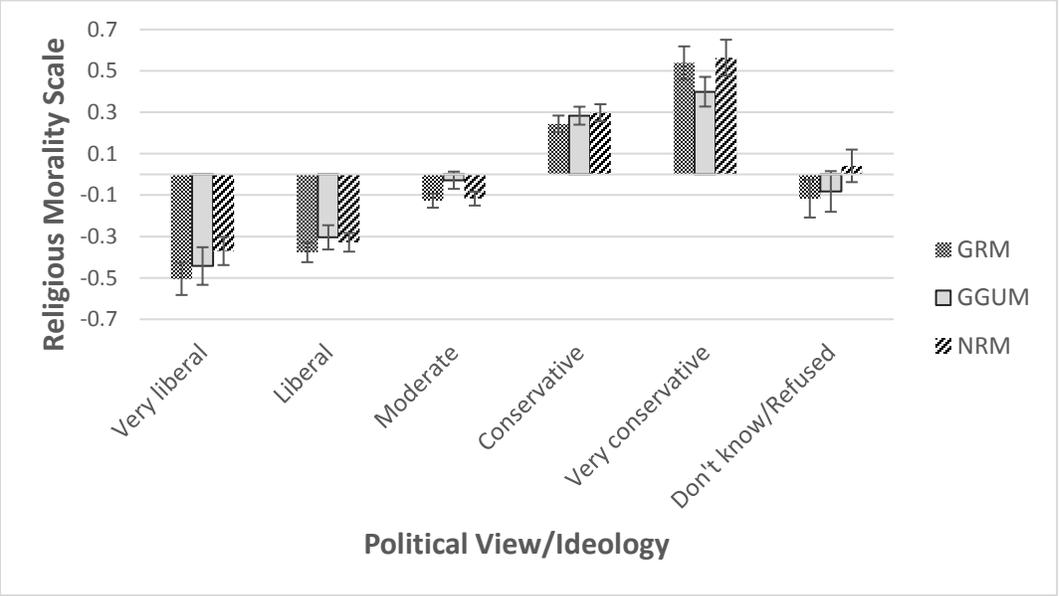

*Figure 5*

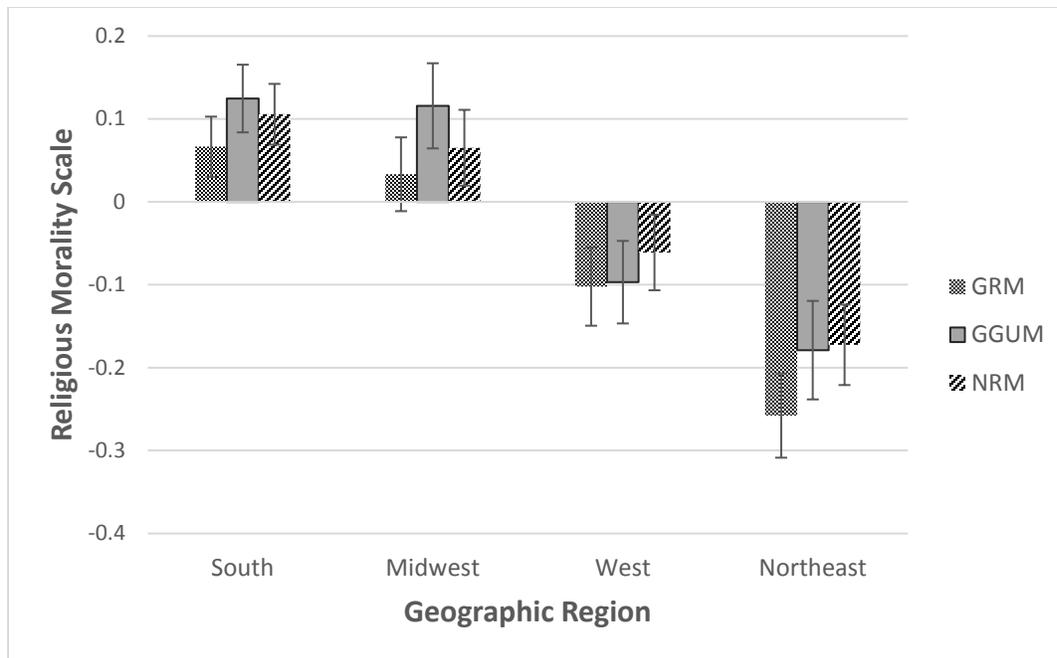

*Figure 6*

## 3.3 Effect of recoding the categories among the IRM approach

Here we check the effects of alternate codings of the categories. As previously discussed, there are situations where the order of the categories might be more subjective. For instance, although the order for ATTEND is straightforward, the order between "Not a moral issue" and "Depend on situation" is open to interpretation.

Another potential coding change is due to fact that codes do not represent magnitude and therefore the gaps between any two codes are not strictly interpretable as true difference. For example, a category that has been coded as 2 is not 2 units away from a category that has been coded as 4. Similarly, the distance between category 1 and 2 cannot be interpreted as mathematically equal to the distance between category 2 and 3.

To compare the effect of revising the order of the response categories, the analyses were conducted on an alternate dataset with the revised codings as shown in Table 5, with changes bolded for emphasis. The "morality" questions were recoded such that *ordering* was changed, whereas the "religiosity" question was revised such that the *distance* between some categories was changed.

| Original ordering | | Revised ordering | |
|---|---|---|---|
| 1 | Morally acceptable | **1** | **Not a moral issue** |
| 2 | Not a moral issue | **2** | **Morally acceptable** |
| 3 | Depend on situation | 3 | Depend on situation |
| 4 | Morally wrong | 4 | Morally wrong |
| NA | Don't know/Refused | NA | Don't know/Refused |
| 6 | More than once a week | **7** | **More than once a week** |
| 5 | Once a week | **6** | **Once a week** |
| 4 | Once or twice a month | 4 | Once or twice a month |
| 3 | A few times a year | 3 | A few times a year |
| 2 | Seldom | 2 | Seldom |
| 1 | Never | **0** | **Never** |
| NA | Don't know/Refused | NA | Don't know/Refused |

*Table 5*

To check whether or not the models changed when the response category codes were changed, model fit metrics were compared. Both AIC and BIC were used and the magnitude of the difference was interpreted using conventions in the literature that differences greater than 10 (i.e., $\Delta_{AIC/BIC} > 10$) indicates strong support that the compared models are in fact different (Burnham & Anderson, 2003; Raftery, 1995). Results show that both GRM and GGUM models are affected by the change in coding. As expected, the NRM model is not affected at all, with the model fit metrics remaining essentially identical between the original and alternate models (Table 6).

| Model | Original model AIC | Alternate model AIC | $\Delta_{AIC}$ | Original model BIC | Alternate model BIC | $\Delta_{BIC}$ |
|---|---|---|---|---|---|---|
| GRM | 14527.03 | 14246.36 | 280.67 | 14622.6 | 14341.92 | 280.68 |
| GGUM | 17211.51 | 14474.68 | 2736.83 | 17328.32 | 14591.49 | 2736.83 |
| NRM | 14199.43 | 14199.37 | 0.06 | 14348.09 | 14348.03 | 0.06 |

*Table 6*

The robustness of the NRM in comparison with the other two models is illustrated in the change in group averages based on the original and alternate coding schemes. Figures 6 through 8 show the effect of recoding on the average scale scores of subgroups by party across the three models. There are observed statistically significant differences in several of the subgroups for both GRM and GGUM. These differences have inferential consequences for the subgroups that are affected because the recoding resulted in average scale scores that are now statistically significantly different from the overall mean (see for example, *Other party* in Figure 7 and *Independent* in Figure 8). It is notable that for NRM, the recoding has no effect at all (Figure 9).

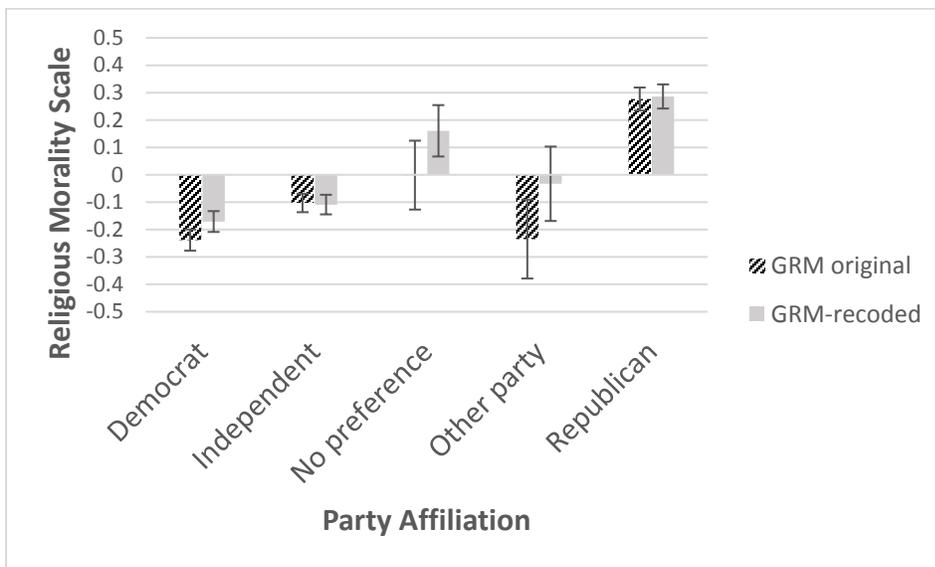

*Figure 7*

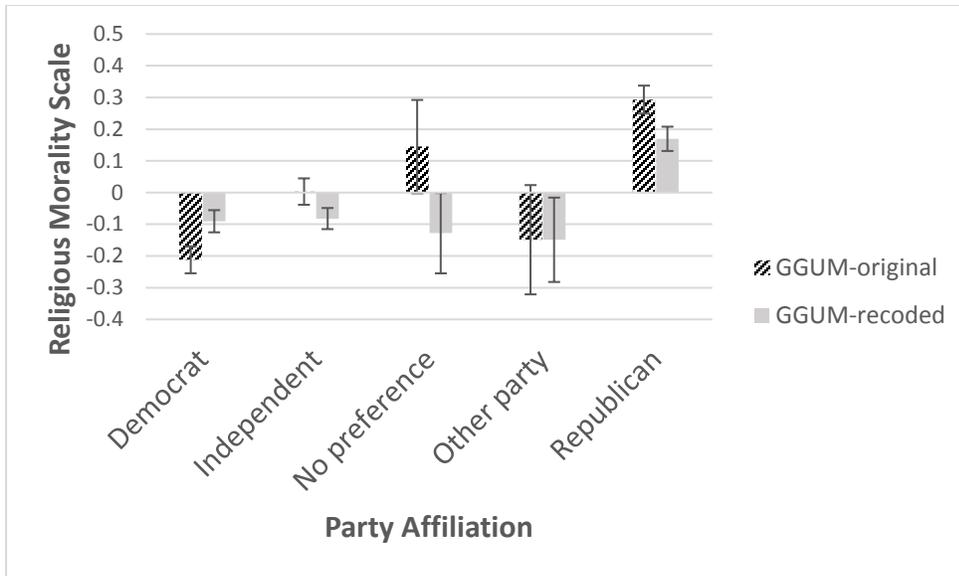

*Figure 8*

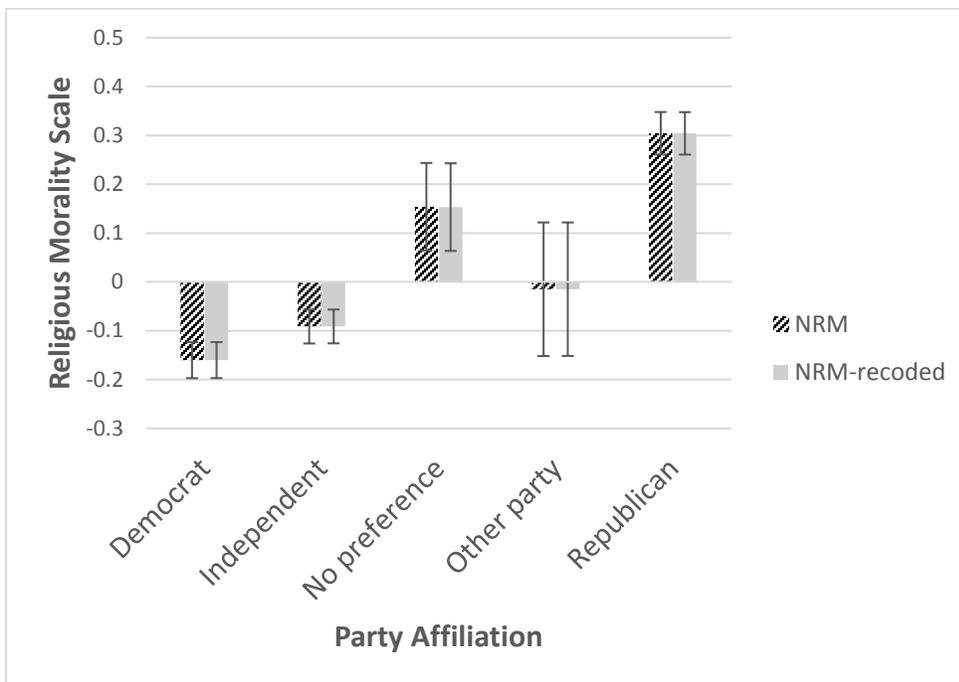

*Figure 9*

Notwithstanding that the NRM is robust to coding changes, the model fit metrics reported in Table 6 suggest that the alternate recoding results in models that are better fitting (lower AIC and BIC values). To enable future researchers to adapt the items in this paper to their instruments, and to allow the results from those to be set on the same scale as what was reported here, the item parameters for the NRM (based on the alternate recoding, Table 5) is reported in Table 7.

| Item | $a_{ik}$ | | | | | | $d_{ik}$ | | | | | |
|---|---|---|---|---|---|---|---|---|---|---|---|---|
| | k=1 | k=2 | k=3 | k=4 | k=5 | k=6 | k=1 | k=2 | k=3 | k=4 | k=5 | k=6 |
| Q40a | -1.285 | -0.519 | 0.196 | 1.608 | | | 1.605 | 1.602 | -1.459 | -1.748 | | |
| Q40B | -1.178 | -0.325 | 0.696 | 0.806 | | | 1.078 | 0.492 | -1.315 | -0.255 | | |
| Q40C | -5.328 | 0.557 | 1.547 | 3.224 | | | -2.355 | 0.66 | 0.279 | 1.416 | | |
| ATTEND | -0.659 | -0.623 | -0.391 | -0.007 | 0.549 | 1.132 | -0.268 | -0.017 | 0.126 | 0.103 | 0.527 | -0.471 |

*Table 7*

## 4 Discussion

The results show that meaningful measurement scales can be developed from very constrained opinion poll data. Utilizing IRM-based approaches can yield measures that are stable under categorical coding changes. More importantly, these measures are interval-level and are thus can be analyzed parametrically. The amount of information that we can extract for inferences on population subgroups is quite substantial given that the data come from only 4 questions obtained through telephone survey. The results based on the NRM estimates show that groups who have combined incomes over $75,000 are significantly different in terms of "religious morality" compared to groups with incomes less than $50,000 (Figure 10). A statistically significant difference is also observable between urban and rural populations (Figure 11). On the other hand, we see no statistically significant differences across various races (Figure 12) and between genders (Figure 13). In addition, all gender and race subgroups are not statistically different from the overall population average (the 95% confidence intervals of all estimates straddles the overall mean).

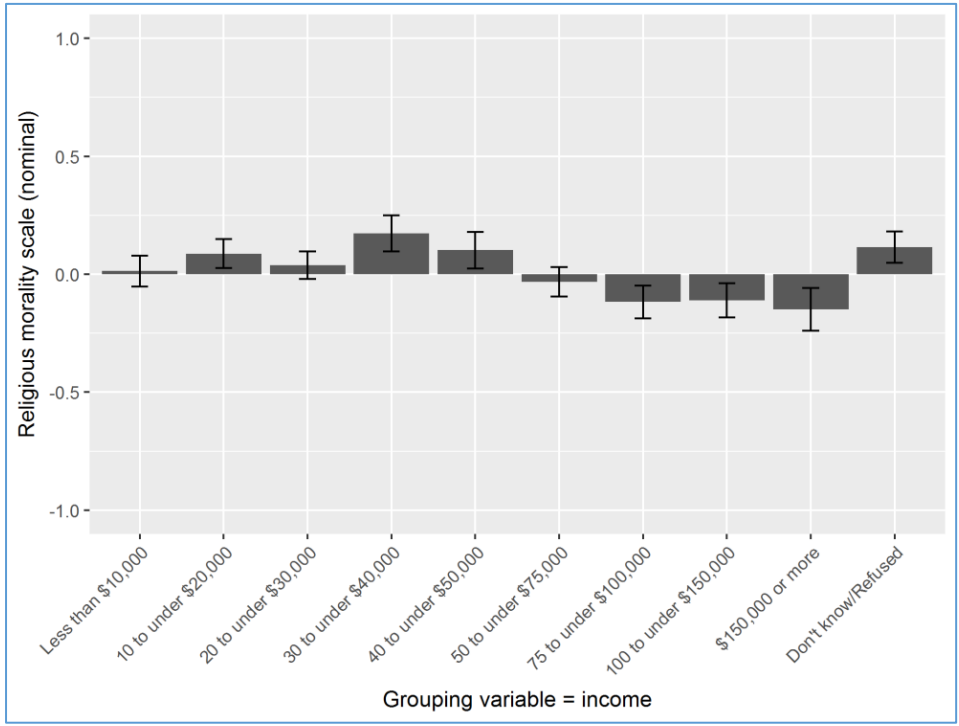

*Figure 10*

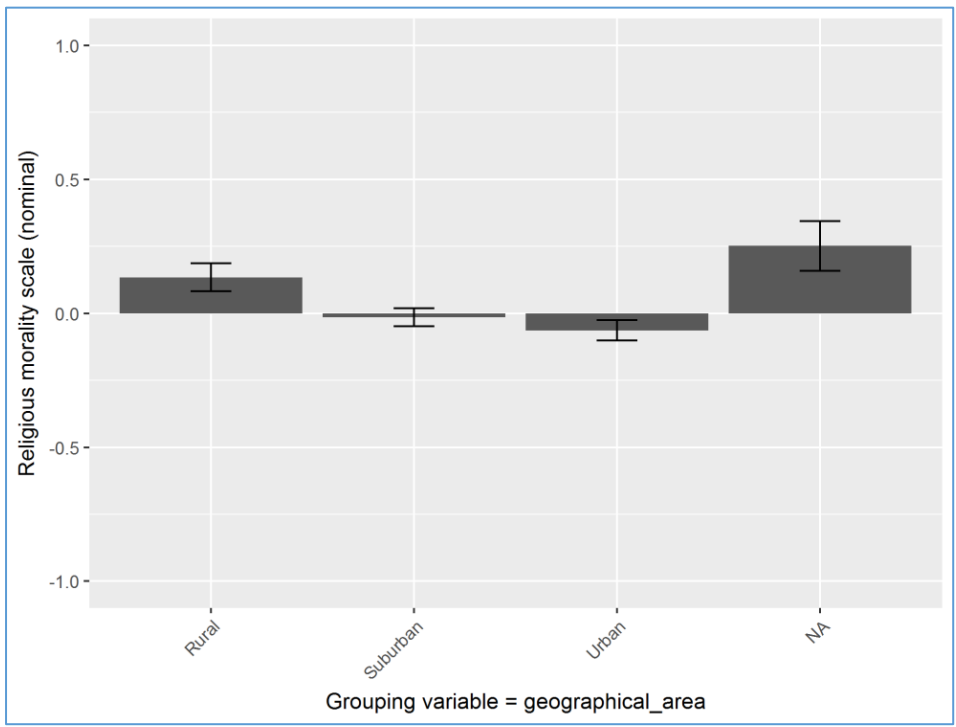

*Figure 11*

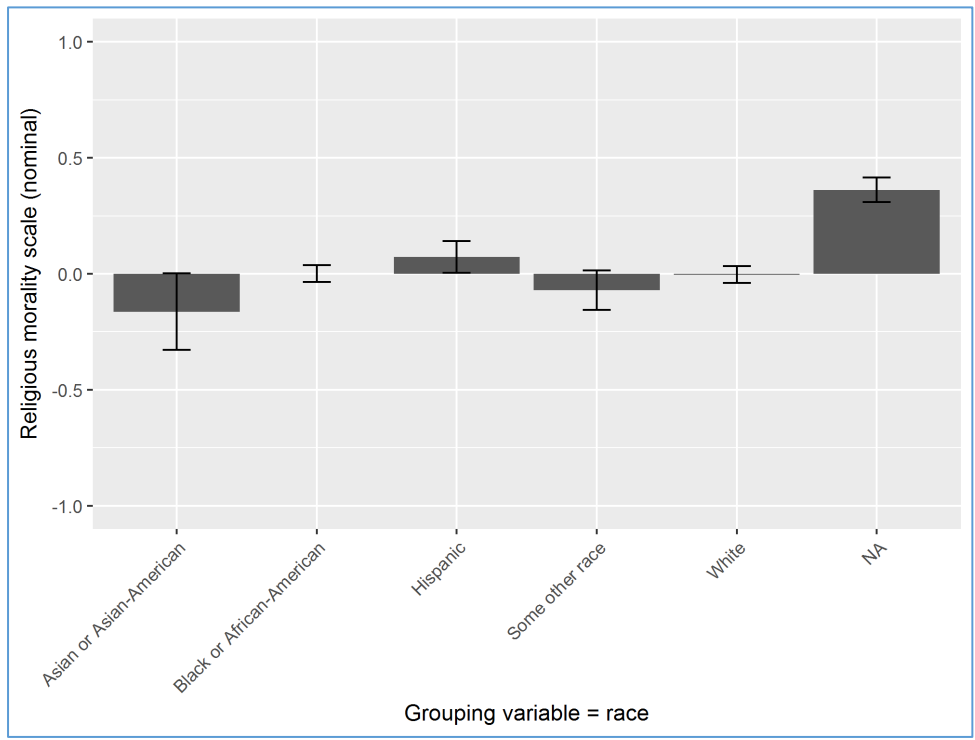

*Figure 12*

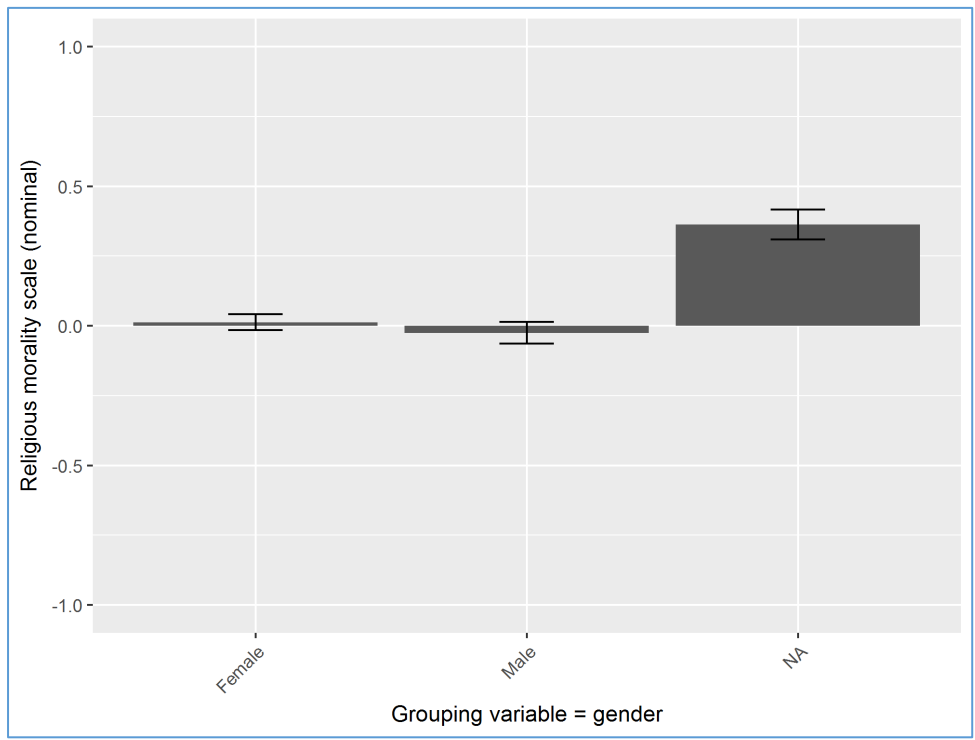

*Figure 13*

Concepts such as religious morality are psycho-social constructs. They are not measurable or even observable directly like variables such as income or educational achievement. However, constructs such as these provide us with important information that is useful for a wide range of political, social, and economic purposes. This makes their measurement an important endeavor. The fact that these constructs are challenging to measure, both in terms of logistics and psychometric issues, highlights the importance of robust measurement approaches.